\documentclass[conference]{IEEEtran}
\IEEEoverridecommandlockouts

\usepackage{cite}
\usepackage{amsmath,amssymb,amsfonts}
\usepackage{algorithmic}
\usepackage{graphicx}
\usepackage{textcomp}
\usepackage{xcolor}
\def\BibTeX{{\rm B\kern-.05em{\sc i\kern-.025em b}\kern-.08em
    T\kern-.1667em\lower.7ex\hbox{E}\kern-.125emX}}

\usepackage{booktabs, array, multirow, multicol}
\usepackage{amsmath, amsfonts}
\usepackage{pifont}
\usepackage{booktabs}
\usepackage{tabularx}
\usepackage{array}

\usepackage{tcolorbox}

\usepackage[svgnames]{xcolor}

\DeclareRobustCommand{\respond_tok}{%
  \textcolor{SteelBlue}{%
    \textbf{\texttt{<|respond|>}}}%
}

\DeclareRobustCommand{\listen_tok}{%
  \textcolor{Salmon}{%
    \textbf{\texttt{<|listen|>}}}%
}

\DeclareRobustCommand{\ignore_tok}{%
  \textcolor{SlateGray}{%
    \textbf{\texttt{<|ignore|>}}}%
}

\begin{document}

\title{Cocktail-Talker: Multi-Speaker Dialog Modeling in Noisy Social Environments with Turn Action GRPO}

\newcommand{\affmark}[1]{\ensuremath{^{\scriptscriptstyle #1}}}

\author{
\IEEEauthorblockN{
Xilin Jiang\affmark{\clubsuit},
Riki Shimizu\affmark{\clubsuit},
Sukru Samet Dindar\affmark{\clubsuit},
Junkai Wu\affmark{\spadesuit},
Zhongweiyang Xu\affmark{\blacklozenge},
Nima Mesgarani\affmark{\clubsuit}
}
\IEEEauthorblockA{
\affmark{\clubsuit}Columbia University,
\affmark{\spadesuit}University of Washington,
\affmark{\blacklozenge}University of Illinois Urbana-Champaign
}
\IEEEauthorblockA{
xj2289@columbia.edu, nima@ee.columbia.edu
}
}
\maketitle

\begin{abstract}
Spoken dialog systems are typically designed for clean, dyadic interactions in which a single user and an assistant take turns speaking. Real-world social conversations, however, are often more ambiguous: multiple speakers may participate in the same conversation amid irrelevant speech and background noise. Each utterance may be directed to the assistant, addressed to another speaker, or completely irrelevant. In such settings, the assistant must decide not only what to say, but also whether to speak at all. In this paper, we introduce \textbf{Cocktail-Talker}, a speech LLM framework for multi-speaker spoken dialog modeling in noisy social environments. We model the assistant's behavior with three action tokens: $\respond_tok{}$, $\listen_tok{}$, and $\ignore_tok{}$, placed before a response or silence. Cocktail-Talker is trained via supervised finetuning and reinforcement learning to generate the appropriate action token and, only in $\respond_tok{}$ mode, a speech response. To prepare the training data, we develop \textbf{Cocktail-DialogGen}, an LLM-based data pipeline that simulates realistic multi-speaker dialogs with speaker roles across diverse social settings. Together, these components take a step toward spoken dialog systems that interact more naturally and selectively in complex social environments. Code: https://github.com/xi-j/Cocktail-Talker
\end{abstract}


\section{Introduction}

Speech conversation presents unique challenges that are largely abstracted away in text conversation. Unlike text dialog, which is written down as separated utterances explicitly associated with their owners, speech reaches a listener's ears as a continuous acoustic signal with no inherent boundaries between turns, speakers, or sound sources. Multiple people can speak spontaneously, often without naming or otherwise identifying themselves; consequently, a listener must infer speaker identity from vocal characteristics and conversational context. Moreover, the incoming signal is often mixed with irrelevant sounds, including other people’s chatter and environmental noise, especially in crowded social settings. These acoustic distractions further compound the ambiguity of multi-speaker conversation, making it difficult to determine both who is speaking and which parts of the scene are relevant.

\begin{figure}[!t]
    \centering
    \includegraphics[width=\linewidth]{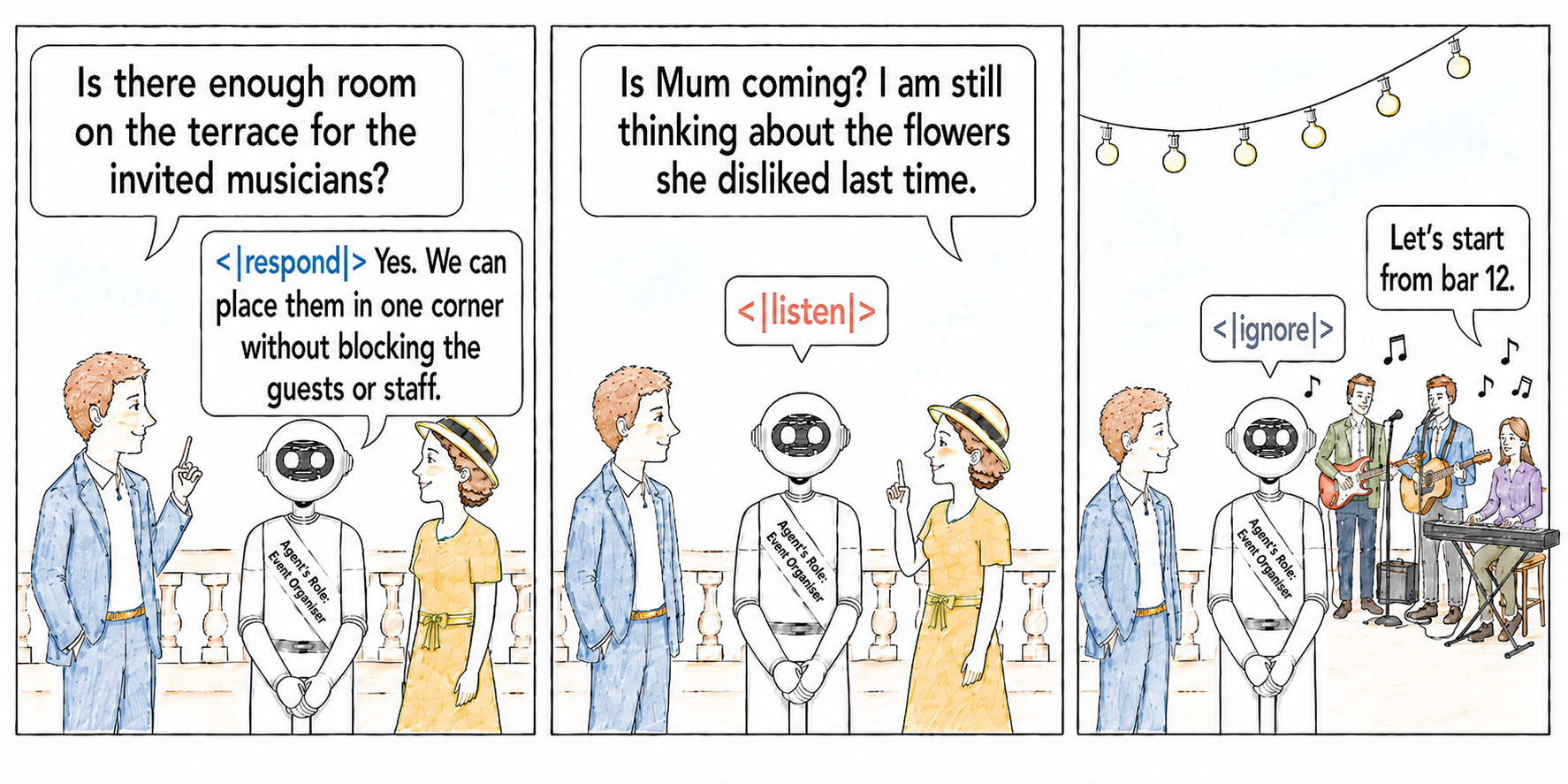}
    \caption{A three-speaker conversation example illustrating Cocktail-Talker's three turn actions. Cocktail-Talker is \textit{text-prompted} to serve as an event organizer. \textbf{Left:} The model \respond_tok{} when the man asks an event-related question that falls within the \textit{Agent's Role}. \textbf{Middle:} The model \listen_tok{} and stays silent when the woman talks to the man rather than the agent. \textbf{Right:} The model \ignore_tok{} for unrelated background speech and music from the rehearsing musicians.}
    \label{fig:concept}
\end{figure}

Despite these challenges, existing spoken dialog systems \cite{gpt4o, moshi, personaplex, qwen25omni, qwen3omni} are typically designed for dyadic interactions between a single user and an assistant in controlled acoustic environments (e.g., a quiet room). The user's speech is directed to the assistant, which is then expected to respond. This assumption does not hold in real-world social environments, where an utterance may be addressed to the assistant, directed toward another person, or irrelevant to the ongoing conversation. A capable speech agent must therefore decide not only what to say, but also whether to respond, continue listening, or ignore the incoming sound altogether, like the example in Fig. \ref{fig:concept}.

In this paper, we study \emph{multi-speaker single-assistant} spoken dialog modeling in noisy social environments, where an assistant is exposed to a continuous mixture of the foreground conversations and the background chatter or noises and must decide how, or whether, to participate. We formulate this setting with three turn actions: $\respond_tok{}$, in which the assistant generates a spoken response; $\listen_tok{}$, in which it remains silent while attending to the ongoing conversation; and $\ignore_tok{}$, in which it disregards speech or sounds that are unrelated to its role or the current interaction. Building on this concept, we introduce \textbf{Cocktail-Talker}, a speech LLM trained with supervised finetuning and reinforcement learning (in particular, GRPO \cite{grpo}) to predict these action tokens and generate a response only when selecting $\respond_tok{}$. To train Cocktail-Talker, we develop \textbf{Cocktail-DialogGen}, a LLM-based, human-in-the-loop data pipeline to simulate realistic multi-speaker conversations across diverse social settings, including indoor and outdoor, working and living spaces. Together, these contribute a framework for building speech assistants or agents that participate selectively and naturally in complex social auditory scenes.

\begin{figure*}
    \centering
    \includegraphics[width=\linewidth]{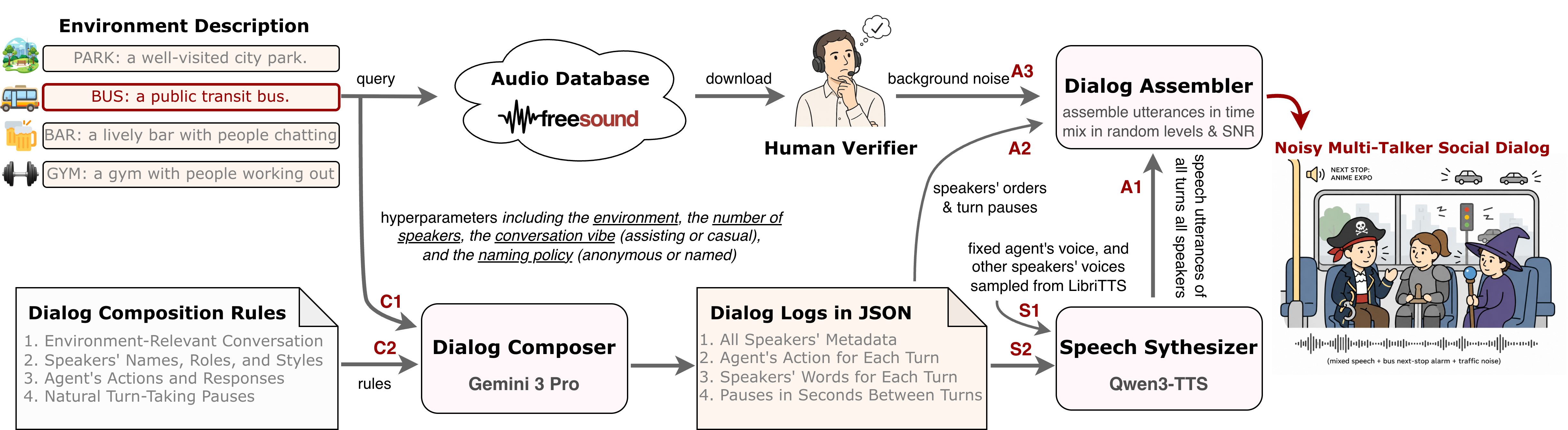}
    \caption{The complete workflow of Cocktail-DialogGen, our simulation pipeline of multi-speaker conversations. Users specify the environment; Dialog composer (Gemini 3 Pro) composes the dialog logs with speakers' metadata and turn actions according to the rules; Qwen3-TTS synthesizes the speech utterances for every speaker; and finally, a dialog assembler program aligns speakers in time with natural time pauses, mixes speech sources with random energy levels, and adds the background audio at a random signal-to-noise ratio (SNR).}
    \label{fig:simulation}
    \vspace{-0.5cm}
\end{figure*}

\section{Related Works}

\noindent
\textbf{Speech Large Language Models.}
Spoken dialog systems have gradually evolved from cascaded pipelines of speech recognition, text generation, and speech synthesis toward more unified, end-to-end architectures, where a large language model (LLM) serves as the backbone for both listening and speaking. Mini-Omni~\cite{miniomni}, LLaMA-Omni~\cite{llamaomni}, and Qwen2-Audio~\cite{qwen2audio} enable low-latency speech-to-speech interaction with LLMs. Qwen2.5-Omni~\cite{qwen25omni} and Qwen3-Omni~\cite{qwen3omni} further extend these capabilities to omni-modal perception and streaming speech generation. Duplex models such as Moshi~\cite{moshi} and Mini-Omni2 \cite{miniomni2} enable real-time interaction by jointly consuming and generating text and speech tokens. PersonaPlex~\cite{personaplex} further builds on Moshi with text- and audio-based system prompts for controlling the agent’s voice and role. While these systems improve the latency and naturalness of dyadic voice interaction, they generally assume that incoming speech is directed to and relevant for the agent.
\\

\noindent
\textbf{Challenges in Multi-Speaker Conversation.} Often referred to as the \textit{``cocktail party problem''} \cite{cherry1953some,bee2008cocktail}, multi-speaker conversation requires the listener (human or model) to process overlapping speech and environmental noise and identify the target speaker and addressee. Despite recent progress in spoken language modeling, several challenges still remain. Study \cite{justasrllm} shows that speech LLMs struggle to infer the target speaker directly from raw audio, while \cite{jiang2026avmeme} finds that models fail to capture conversation context from audio alone. \text{MSU-Bench} \cite{msubench} reports degraded performance in multi-speaker conversation understanding as complexity increases, \text{M3-SLU} \cite{M3-SLU} exposes model failures in speaker attribute reasoning, and \text{Full-Duplex-Bench} \cite{fullDuplexBench,lin2026full} reveals fragile model behaviors under turn-taking dynamics and background speech. Nevertheless, these works primarily discover and explain these limitations rather than develop practical solutions for robust multi-speaker interaction.
\\

\noindent
\textbf{Solutions to Multi-Speaker Conversation.} Existing models and systems mostly focus on the \textit{listening} side of multi-speaker conversation, including target speaker extraction \cite{voicefilter, lcr, shimizu2025meanflow}, speech transcription \cite{kanda2019aux,denisov2019speakeremb, zhang2023tsasr}, and conversation understanding \cite{aadllm, yin2026focus}, on synthetic or real datasets \cite{librimix, chime6, amimeeting}. The \textit{speaking} side, \textit{whether and how to respond}, remains under-explored. A conceptually similar work \cite{bhagtani2026speak} studies the response decision in textual (not spoken) conversation, without voice identity, timing, and noise. In this work, we propose Cocktail-Talker to bridge this gap with the concept of a spoken dialog assistant.

\section{Cocktail-DialogGen}
\label{cocktail-dialoggen}

Before introducing the model, it is necessary to first characterize the noisy multi-speaker social conversations considered in this work. In our \textit{cocktail party} setting, an assistant receives a continuous audio mixture comprising: (1) \textbf{a target conversation}, in which \underline{three} or \underline{four} speakers take turns talking; and (2) \textbf{background sounds}, including speech from non-target conversations, sound effects, and ambient noise. Despite the abundance in the real world, collecting at scale with clean speech sources and accurate annotations for the assistant is difficult. Therefore, we adopt a simulation-based approach that constructs cocktail-party conversations from the bottom up.

\begin{table}[!t]
\centering
\caption{A summary of Input arguments and \\ fixed constraints of Cocktail-DialogGen.}
\label{tab:dialoggen_params}
\footnotesize
\setlength{\tabcolsep}{2pt}
\renewcommand{\arraystretch}{1.03}
\begin{tabular}{p{0.29\columnwidth} p{0.64\columnwidth}}
\hline
\multicolumn{2}{c}{\textbf{Input Arguments}} \ \\
\hline
Number of Speakers & 3 or 4 \\
Environment Category & 18 seen / 10 unseen \\
Conversation Vibe & \textsc{ASSISTING} (formal, professional) \newline
\textsc{CASUAL} (acquaintance, relaxed) \\
Naming Policy & \textsc{Named}: allow names \& roles called \newline \textsc{Anonymous}: addressee inferred from context \\
\hline
\multicolumn{2}{c}{\textbf{Fixed Constraints}} \\
\hline
\multicolumn{2}{c}{For Dialog Composer (Gemini 3 Pro)} \\
\hline
Thinking Level & Medium \\
Required Actions & All three actions: respond, listen, and ignore \\
Addressee Switches & At least once \\
\hline
\multicolumn{2}{c}{For Speech Synthesizer (Qwen3-TTS)} \\
\hline
Assistant's Voice & Fixed to Qwen2.5-Omni's ``Chelsie'' \\
Other Speakers & LibriTTS train, dev, test \\
\hline
\multicolumn{2}{c}{For Dialog Assembler} \\
\hline
Assistant's Energy & $-20\mathrm{dB}$ RMS \\
Other Spks' Energies & $-20 \pm 5\mathrm{dB}$ RMS \\
Conversation SNR & Five ranges: \{0--3, 3--6, 6--9, 9--12\}$\mathrm{dB}$ and clean\\
\hline
\label{tab:data_config}
\end{tabular}
\vspace{-0.8cm}
\end{table}

Fig. \ref{fig:simulation} shows the workflow of \textbf{Cocktail-DialogGen}, our multi-speaker conversation simulation pipeline. Cocktail-DialogGen is \textit{environment}-oriented: it begins with a natural-language description of an environment, such as \textcolor{gray}{``PARK: a well-visited city park''}. This description steers both the acoustic and semantic aspects of the simulation: the selection of background sounds and the generation of conversations that are contextually consistent with the environment. In principle, users may provide either a free-form environment description or an environment recording, which can first be captioned by an audio large language model. However, to ensure data quality and to evaluate model generalization, we simulate and train on 18 fixed environment categories defined by the DEMAND \cite{DEMAND} noise database, including indoor (e.g., kitchen and living room), outdoor (e.g., sports field and river), and transportation (e.g., car and metro) settings. We propose 10 unseen environments (e.g., airport, zoo, and ferry) to test the model's generalization. While we borrow the environment categories from DEMAND, it contains too few recordings to train the model. Therefore, we downloaded by relevance and verified by humans 100 recordings for each environment category from Freesound (https://freesound.org/). Except for the \textcolor{gray}{``MEETING''} category, we used 100 DailyTalk \cite{dailytalk} samples as the background (non-target) conversations. We reserved the original noise recordings from DEMAND to construct the test set of seen environments instead.

The remaining input arguments in Table~\ref{tab:data_config} explicitly prompt Gemini 3 Pro \cite{gemini3pro}, our \textbf{Dialog Composer}, to generate dialogs with a specified number of speakers, conversation vibe, and naming policy. Crucially, since explicit name or role mentions substantially simplify addressee identification for the assistant to decide whether to respond or not, the \textsc{Named} setting permits speakers to address one another using their names (\textit{e.g., ``Alex''}) or roles  (e.g., \textit{``Officer''} or \textit{``Daughter''}), but the \textsc{Anonymous} setting prohibits such references in any sentence of the dialog and requires the addressee to be inferred from conversational context. We post-process the dialogs to enforce this constraint. 

\begin{figure}[!t]
\centering
\newcommand{\I}{\phantom{xx}}%
\begin{tcolorbox}[
  colback=Wheat!15, colframe=Wheat!15,
  boxrule=0pt, arc=2pt,
  left=2pt, right=2pt, top=1pt, bottom=1pt,
  width=0.9\columnwidth
]
\scriptsize\ttfamily
\textbf{[Metro / Assisting / Anonymous / 3-Spk]}\\[2pt]
"environment": "METRO: a subway"\\
"activity": "Navigating to a city landmark"\\
"voice\_profiles": \{\\
\I "agent": \{\\
\I\I gender: female, \\
\I\I name: Rachel,\\
\I\I role: Station Manager,\\
\I\I style: professional, calm, articulate\\
\I \},\\
\I "speaker\_b": \{\\
\I\I gender: male, \\
\I\I name: Tom,\\
\I\I role: Lost Tourist,\\
\I\I style: anxious, speaks fast\\
\I \},\\
\I "speaker\_c": \{\\
\I\I gender: female, \\
\I\I name: Lisa,\\
\I\I role: Local Guide,\\
\I\I style: confident, slightly impatient\\
\I \}\\
\I "speaker\_d": ...\\
\}\\[4pt]
\textrm{--- Turn 1 ---}\\
action: $\respond_tok{}$,\ pause: 1.2s\\
speaker\_b:\ "Excuse me, we are trying to reach the museum district. Are we on the right platform?"\\
agent: "You are on the southbound platform. To reach the museum district, take the northbound line."\\[4pt]
\textrm{--- Turn 2 ---}\\
action: $\listen_tok{}$,\ pause: 0.4s\\
speaker\_c:\ "See? I told you we swiped in on the wrong side."\\
agent: \textit{<SILENCE>}\\
\textrm{--- More Turns Skipped ---}
\end{tcolorbox}
\caption{An example three-speaker dialog with speakers' metadata, content and action for each turn, and between-turn pauses.}
\label{fig:dialog_example}
\vspace{-0.5cm}
\end{figure}

Each Gemini call generates 12 unique conversations that can happen in the specified environment, as diverse as possible. Fig.~\ref{fig:dialog_example} shows a snapshot of an example dialog log, structured from the agent's (a.k.a. assistant, used interchangeably in the paper) perspective, with all content and annotations relative to the agent. Each log begins with a metadata block that is generated prior to the dialog content: the agent's and other speakers' \textit{\textbf{genders}}, \textit{\textbf{names}}, and \textit{\textbf{roles}} are optionally included in the model's text input, reflecting the realistic assumption that the agent knows herself and the people she is conversing with; whereas the environment category, activity, and speaking styles are used only to condition the dialog generation and are withheld from the model. Beyond the speech transcripts, each turn is annotated with an action label and a pause duration. Various pause durations add dynamics to the conversation: short pauses (0–0.5 s) are used for fluid, reactive exchanges, whereas longer pauses (2–5 s) are used for topic changes, distractions, or intervals containing only background sound, the latter being the use case of the $\ignore_tok{}$ action. 

Given the dialog logs, Qwen3-TTS~\cite{qwen3tts}, the \textbf{Speech Synthesizer}, converts each utterance transcript into a speech waveform. The agent's voice is fixed to ``Chelsie'', the default female voice of Qwen2.5-Omni \cite{qwen25omni}. The other speakers' voices are randomly sampled from LibriTTS~\cite{libritts}, giving each dialog a distinct cast of voices. To suppress TTS failures, any utterance exceeding 1.5 seconds per word is rejected and regenerated up to 5 times. Finally, the \textbf{Dialog Assembler} program assembles the per-utterance waveforms into a continuous audio mixture, placing each utterance according to the pause durations specified in the dialog log and normalizing speaker energies as listed in Table~\ref{tab:data_config}. For each unique dialog, we mix four environment recordings (from Freesound or DailyTalk) at four SNR levels, besides the clean version without background, where SNR is measured relative to the active speech RMS to avoid silence gaps diluting the noise level.

In total, we prepared 14,400 unique dialogs for training ($18~\text{environments} \times 2~\text{vibes} \times 2~\text{naming policies} \times 2~\text{speaker counts} \times 100~\text{dialogs}$), each mixed at five noise levels to yield 72,000 noisy audio mixtures (${\approx}1{,}280$ hours). For evaluation, we generated 10 dialogs per condition, yielding 1440 unique dialogs / 7200 noisy mixtures for seen and 800 unique dialogs / 4000 noisy mixtures for unseen environments.

\section{Cocktail-Talker}

Cocktail-Talker is built upon Qwen2.5-Omni-7B~\cite{qwen25omni}, a speech LLM with a \textit{thinker--talker} architecture: the \textit{thinker} is a transformer language model that processes audio input and generates text responses, and the \textit{talker} is a streaming TTS decoder that synthesizes those responses into speech. While the idea can be instantiated in other speech LLMs as well, we choose Qwen2.5-Omni for its strong spoken dialog performance at a moderate model size, enabling efficient training and inference on consumer-grade hardware.
\noindent\rule{\linewidth}{0.4pt}

\noindent\textbf{Input and Output Format.}
At each dialog turn, the model receives a text prompt and an audio input. The text prompt supplies the agent's optional speakers' metadata:

\begin{tcolorbox}[
  colback=Wheat!15, colframe=Wheat!15,
  boxrule=0pt, arc=1pt,
  left=1pt, right=1pt, top=3pt, bottom=1pt
]
\setlength{\leftmargini}{0.5em}
\begin{quote}
\scriptsize\ttfamily
\newcommand{\I}{\phantom{xx}}%
\newcommand{\PH}[1]{$\langle$\textrm{\textit{#1}}$\rangle$}%
Speakers' Metadata: \{\\
\I 'voice\_profiles': \{\\
\I\I 'agent':\ \ \ \ \{\\
\I\I\I gender: \PH{gender},\ name: \PH{name},\ role: \PH{role}\\
\I\I \},\\
\I\I 'speaker\_b': \{\\
\I\I\I gender: \PH{gender},\ name: \PH{name},\ role: \PH{role}\\
\I\I \},\\
\I\I 'speaker\_c': \{\\
\I\I\I gender: \PH{gender},\ name: \PH{name},\ role: \PH{role}\\
\I\I \}\\
\I \},\\
\I 'mode': \PH{assisting $|$ casual}\\
\}\\
$\langle$audio$\rangle$ Generate the next response of the agent speaker.
\end{quote}
\end{tcolorbox}

\noindent and the $\langle$audio$\rangle$ input is the \textit{prefix clip}: the full noisy mixture of all preceding turns, without speech separation, denoising, or chunking, trimmed to the end of the current input utterance. The model (agent) outputs one of three action tokens followed by a spoken response only for $\respond_tok{}$:

\vspace{2pt}
\begin{tcolorbox}[
  colback=Wheat!15, colframe=Wheat!15,
  boxrule=0pt, arc=2pt,
  left=4pt, right=4pt, top=3pt, bottom=3pt
]
\scriptsize\ttfamily
\begin{tabular}{@{}c@{\,}ll}
\multirow{3}{*}{$\Bigg\{$}
  & \quad \respond_tok{} \textit{$\langle$response text$\rangle$} & agent speaks \\
  & \quad \listen_tok{} & agent stays silent \\
  & \quad \ignore_tok{} & agent stays silent \\
\end{tabular}
\end{tcolorbox}
\vspace{2pt}

\noindent The action tokens are skipped when feeding the thinker's output to the talker, so that the talker only synthesizes the response text, preserving Qwen2.5-Omni's original speech quality. To overcome \textbf{\textit{partial or missing metadata}} in inference, we randomly drop each speaker attribute {$\langle$\textrm{\textit{gender}}$\rangle$}, {$\langle$\textrm{\textit{name}}$\rangle$}, {$\langle$\textrm{\textit{role}}$\rangle$}, {$\langle$\textrm{\textit{assisting $|$ casual}}$\rangle$} with probability $p = 0.5$ in training.

\noindent\rule{\linewidth}{0.4pt}

\noindent\textbf{Trainable Components.}
We finetune the thinker using LoRA~\cite{lora} with rank $r{=}128$ and $\alpha{=}256$, applied to all linear layers across all 28 transformer layers. Three special tokens $\respond_tok{}$, $\listen_tok{}$, $\ignore_tok{}$ are newly added to the vocabulary and only their rows in the embedding table and language model head are made trainable. All other embedding weights are frozen. The audio encoder and the talker are kept frozen throughout.

\noindent\rule{\linewidth}{0.4pt}


\begin{table}[t]
\centering
\setlength{\tabcolsep}{2.5pt}
\renewcommand{\arraystretch}{1.12}
\resizebox{\columnwidth}{!}{%
\begin{tabular}{>{\raggedright\arraybackslash}p{3.0cm}|cccc|cccc}
\toprule
\multirow{2}{*}{\textbf{Model}}
  & \multicolumn{4}{c|}{\texttt{SEEN}}
  & \multicolumn{4}{c}{\texttt{UNSEEN}} \\
\cmidrule(lr){2-5}\cmidrule(lr){6-9}
  & Acc & F1-R & F1-S & F1-mac
  & Acc & F1-R & F1-S & F1-mac \\
\midrule
Moshi$^\dagger$
  & 0.424 & 0.596 & 0.000 & 0.298
  & 0.428 & 0.599 & 0.000 & 0.300 \\
PersonaPlex$^\dagger$
  & 0.424 & 0.596 & 0.000 & 0.298
  & 0.428 & 0.599 & 0.000 & 0.300 \\
\midrule
Step-Audio2
  & 0.454 & 0.536 & 0.337 & 0.437
  & 0.470 & 0.542 & 0.370 & 0.456 \\
Kimi-Audio
  & 0.596 & 0.226 & 0.726 & 0.476
  & 0.605 & 0.320 & 0.722 & 0.521 \\
Qwen2.5-Omni
  & 0.512 & 0.502 & 0.522 & 0.512
  & 0.504 & 0.518 & 0.490 & 0.504 \\
Qwen3-Omni
  & 0.600 & 0.255 & 0.727 & 0.491
  & 0.601 & 0.271 & 0.725 & 0.498 \\
\midrule
\textit{w/o action tokens}
  & 0.907 & 0.878 & 0.924 & 0.901
  & 0.908 & 0.882 & 0.925 & 0.903 \\
\textbf{Ours} (SFT)
  & 0.908 & 0.881 & 0.926 & 0.903
  & 0.912 & 0.887 & 0.928 & 0.907 \\
\textbf{Ours} (SFT+GRPO)
  & \textbf{0.931} & \textbf{0.914} & \textbf{0.942} & \textbf{0.928}
  & \textbf{0.933} & \textbf{0.917} & \textbf{0.943} & \textbf{0.930} \\
\bottomrule
\end{tabular}
}
\caption{Turn decision accuracy. All models are evaluated on binary
Respond/Silent (R/S) accuracy (\textbf{Acc}) and per-class F1. $^\dagger$Speech-to-speech
models without a silent mechanism.}
\label{tab:decision_accuracy}
\vspace{-0.5cm}
\end{table}

\begin{figure}
    \centering
    \includegraphics[width=0.9\linewidth]{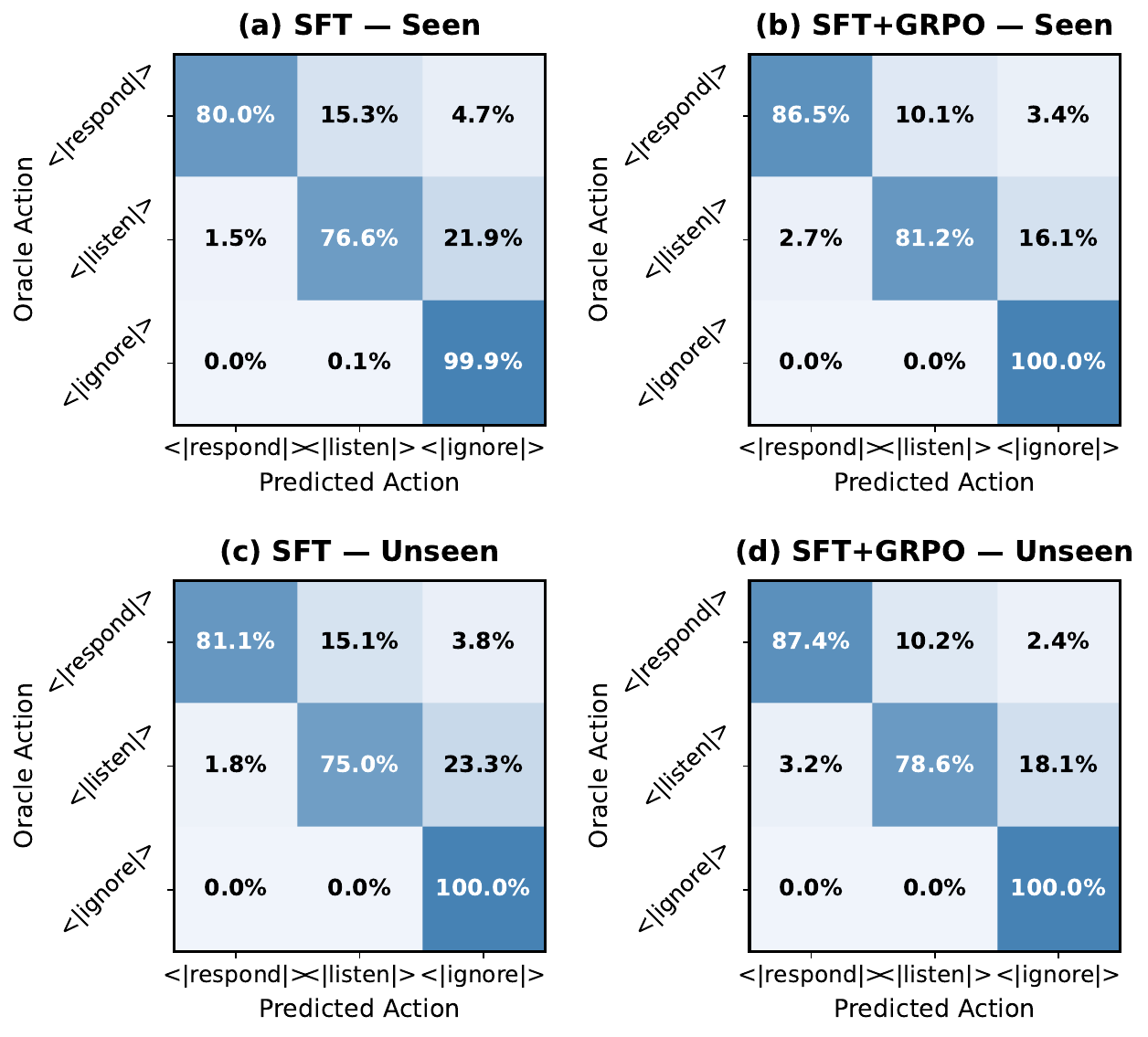}
    \caption{Cocktail-Talker's confusion matrices on seen (a,b)
    and unseen (c,d) environments before and after GRPO. GRPO primarily
    recovers \respond_tok{} recall while reducing \listen_tok{}/\ignore_tok{}
    confusion.}
    \label{fig:confusion}
\end{figure}

\noindent\textbf{Supervised Finetuning (SFT).}
We train on all 580,100 turns from the 72,000 conversation mixtures simulated by Cocktail-DialogGen for one epoch using LlamaFactory~\cite{llamafactory}. We use a per-device batch size of 3 across 4 NVIDIA L40 GPUs, a learning rate of $4{\times}10^{-5}$ with a cosine schedule and 10\% warmup, and mixed-precision bfloat16. Training runs for 48,342 steps and completes in approximately 52 hours.

\noindent\rule{\linewidth}{0.4pt}

\begin{table*}[t]
\centering
\small
\setlength{\tabcolsep}{3pt}
\renewcommand{\arraystretch}{1.12}
\resizebox{\textwidth}{!}{%
\begin{tabular}{>{\raggedright\arraybackslash}p{4.1cm}|cccc|cccc}
\toprule
\multirow{2}{*}{\textbf{Model}}
  & \multicolumn{4}{c|}{\texttt{SEEN ENVIRONMENTS}}
  & \multicolumn{4}{c}{\texttt{UNSEEN ENVIRONMENTS}} \\
\cmidrule(lr){2-5}\cmidrule(lr){6-9}
  & METEOR & ROUGE-L & BERTScore & SentCos
  & METEOR & ROUGE-L & BERTScore & SentCos \\
\midrule
Moshi$^\dagger$
  & 0.071 & 0.077 & 0.847 & 0.172
  & 0.077 & 0.078 & 0.844 & 0.174 \\
PersonaPlex$^\dagger$
  & 0.115 & 0.087 & 0.838 & 0.264
  & 0.120 & 0.095 & 0.839 & 0.284 \\
\midrule
Step-Audio2
  & 0.093\,/\,0.125 & 0.074\,/\,0.100 & 0.627\,/\,0.844 & 0.241\,/\,0.324
  & 0.086\,/\,0.117 & 0.074\,/\,0.100 & 0.620\,/\,0.845 & 0.238\,/\,0.325 \\
Kimi-Audio
  & 0.017\,/\,0.121 & 0.023\,/\,0.165 & 0.121\,/\,0.868 & 0.051\,/\,0.364
  & 0.028\,/\,0.128 & 0.038\,/\,0.175 & 0.189\,/\,0.869 & 0.086\,/\,0.395 \\
Qwen2.5-Omni
  & 0.077\,/\,0.132 & 0.080\,/\,0.139 & 0.502\,/\,0.867 & 0.212\,/\,0.366
  & 0.084\,/\,0.135 & 0.086\,/\,0.138 & 0.539\,/\,0.866 & 0.236\,/\,0.380 \\
Qwen3-Omni
  & 0.021\,/\,0.128 & 0.025\,/\,0.152 & 0.139\,/\,0.862 & 0.057\,/\,0.351
  & 0.025\,/\,0.146 & 0.027\,/\,0.157 & 0.150\,/\,0.867 & 0.065\,/\,0.377 \\
\midrule
\textit{(SFT without action tokens)}
  & 0.175\,/\,0.221 & 0.219\,/\,0.276 & 0.713\,/\,0.899 & 0.385\,/\,0.485
  & 0.159\,/\,0.197 & 0.201\,/\,0.250 & 0.720\,/\,0.893 & 0.373\,/\,0.463 \\
\textbf{Cocktail-Talker} (SFT)
  & 0.183\,/\,\underline{0.229} & 0.225\,/\,\underline{0.282} & 0.719\,/\,\underline{0.900} & 0.391\,/\,\underline{0.489}
  & 0.161\,/\,\underline{0.199} & 0.203\,/\,\underline{0.250} & 0.725\,/\,\underline{0.893} & 0.381\,/\,\underline{0.470} \\
\textbf{Cocktail-Talker} (SFT+GRPO)
  & \textbf{0.194}\,/\,0.225 & \textbf{0.240}\,/\,0.278 & \textbf{0.777}\,/\,0.899 & \textbf{0.418}\,/\,0.484
  & \textbf{0.171}\,/\,0.195 & \textbf{0.216}\,/\,0.247 & \textbf{0.781}\,/\,0.893 & \textbf{0.407}\,/\,0.465 \\
\bottomrule
\end{tabular}
}
\caption{Response quality. Each cell reports
\textit{penalized\,/\,conditional} scores: the penalized variant averages
over all oracle-respond turns (silent prediction scores 0); the conditional
variant averages only over turns where both oracle and model produced
text.}
\label{tab:response_quality}
\end{table*}

\noindent\textbf{Group Relative Policy Optimization (GRPO).}
Starting from the merged SFT checkpoint, we apply GRPO~\cite{grpo} using ms-swift~\cite{ms-swift} with vLLM~\cite{vllm} to accelerate rollout generation. For each training prompt $q$, $G{=}16$ completions $\{o_i\}_{i=1}^{G}$ are sampled and scored. The group-normalized advantage is
\begin{equation}
    \hat{A}_i = \frac{r_i - \mathrm{mean}(\{r_j\})}{\mathrm{std}(\{r_j\})}
\end{equation}
and the training objective $\mathcal{L} =$
\begin{equation}
\hspace{-0.2cm}
     -\frac{1}{G}\sum_{i=1}^{G}\Bigl[\min\!\left(\rho_i\hat{A}_i,\,\mathrm{clip}(\rho_i,1{\pm}\epsilon)\hat{A}_i\right) - \beta\,D_\mathrm{KL}(\pi_\theta\|\pi_\mathrm{ref})\Bigr]
\end{equation}
where $\rho_i = \pi_\theta(o_i|q)/\pi_{\theta_\mathrm{old}}(o_i|q)$. The reward $r_i \in [0, 2]$ is the sum of two components: (1) \textbf{Action accuracy} $\in [0,1]$: 1.0 for a correct action and 0.0 for incorrect action; (2) \textbf{Format integrity} $\in [0,1]$: 1.0 if the output begins with exactly one action token and obeys the structural rules ($\listen_tok{}/\ignore_tok{}$ followed by nothing; $\respond_tok{}$ followed by non-empty text). 

We run GRPO for 2,000 steps with a learning rate of $1{\times}10^{-5}$ (cosine, 5\% warmup), per-device batch size 2, gradient accumulation 2, across 4 NVIDIA L40 GPUs in bfloat16. Training completes in approximately 4.5 hours.

\begin{figure*}[!th]
    \centering
    \includegraphics[width=\textwidth]{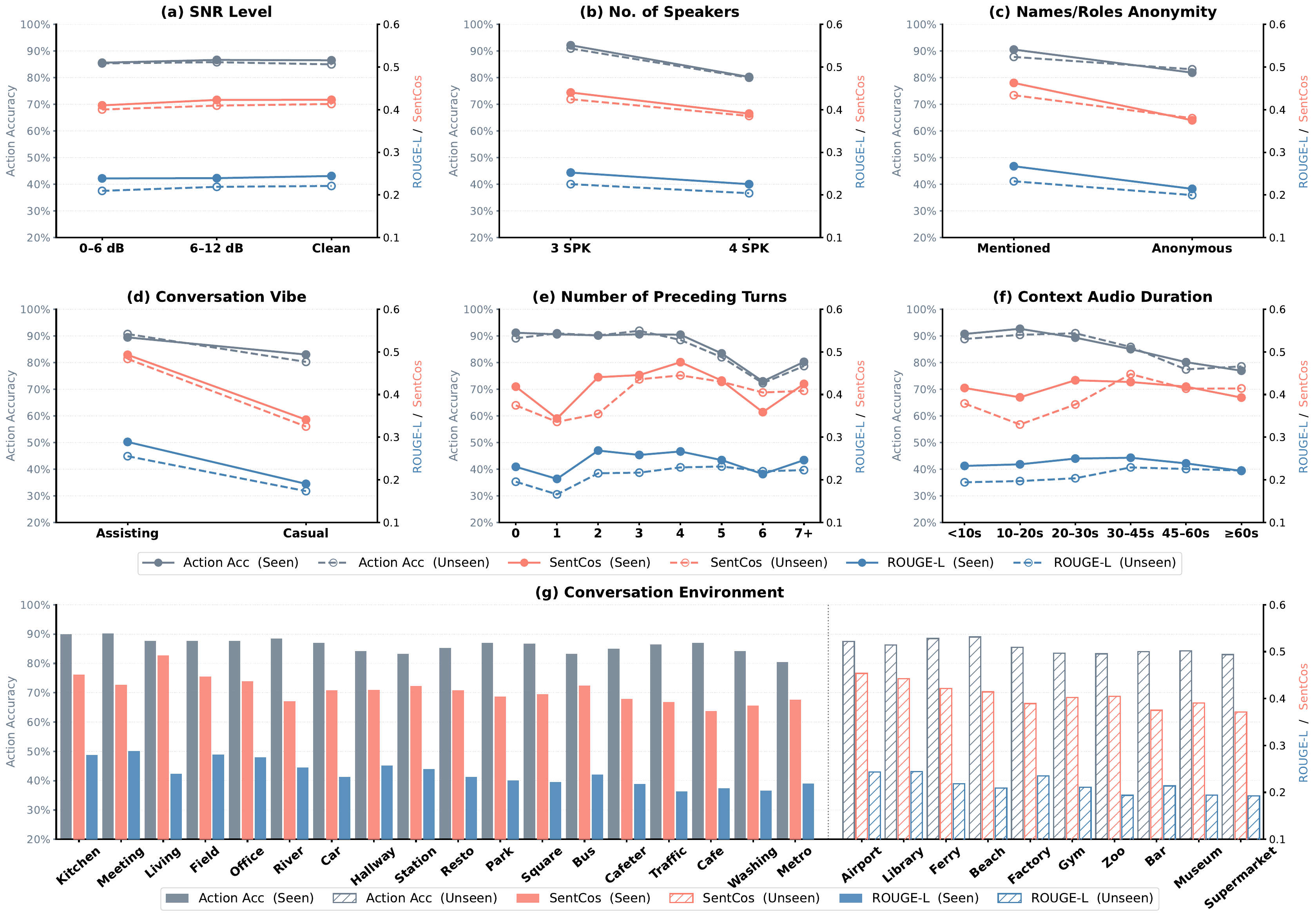}
    \caption{Cocktail-Talker (SFT+GRPO) performance by (a) SNR level, (b) number of
    speakers, (c) names/roles anonymity, (d) conversation vibe, (e) number of preceding turns, (f) context audio duration, and (g) environment type. Solid/dashed lines and filled/hatched bars denote seen/unseen
    splits respectively.}
    \label{fig:analysis}
\end{figure*}

\section{Results and Analysis}

\noindent\textbf{Evaluation Protocol.} We compare against strong speech LLMs. Moshi~\cite{moshi} and PersonaPlex~\cite{personaplex} are speech-to-speech models without text instruction interfaces, whereas Step-Audio2~\cite{stepaudio2}, Kimi-Audio~\cite{kimiaudio}, Qwen2.5-Omni~\cite{qwen25omni}, and Qwen3-Omni~\cite{qwen3omni} accept text input. The latter four receive all speakers' metadata in natural language and are instructed to decide whether to speak or remain silent based on speaker identities and conversational context. PersonaPlex receives the agent's metadata as a natural-language system prompt. For each model, we construct matched test variants by replacing the agent's voice with that model's synthesized voice, allowing it to condition on its own voice. Because no baseline distinguishes \textit{listen} from \textit{ignore}, we collapse them into a binary Respond/Silent (R/S) decision in Table~\ref{tab:decision_accuracy}: \textbf{Acc} denotes binary R/S accuracy, F1-R and F1-S are per-class F1 scores, and F1-macro is their average. For response quality (Table~\ref{tab:response_quality}), we compute METEOR~\cite{meteor}, ROUGE-L~\cite{rouge}, BERTScore~\cite{bertscore}, and SentCos (cosine similarity between Sentence-BERT~\cite{sentence-bert} embeddings) between oracle and predicted responses. Each metric is reported as \textit{X\,/\,Y}: the \textit{penalized} score $X$ averages over all oracle-respond turns, assigning zero to missed responses, whereas the \textit{conditional} score $Y$ averages only turns where both the oracle and model produce text, measuring response quality conditional on responding. For Moshi and PersonaPlex, which always produce speech, the penalized and conditional scores are identical; we therefore report a single value.

\noindent\rule{\linewidth}{0.4pt}

\noindent\textbf{Decision Accuracy (Table~\ref{tab:decision_accuracy}).}
Speech-to-speech models without instructional text prompts and trained without multi-speaker conversations (Moshi and PersonaPlex) always produce speech, achieving F1-S\,$=$\,0 and near-chance macro F1.
Prompted models (Qwen2.5-Omni, Step-Audio2, Kimi-Audio, and Qwen3-Omni) improve somewhat but remain below 0.6 macro F1, with notably unbalanced F1-R/F1-S. Cocktail-Talker (SFT+GRPO) achieves the best decision accuracy across both splits (0.928/0.930 macro F1), primarily by recovering missed responses: Fig.~\ref{fig:confusion} shows respond recall rising from 80\% to 87\% (seen) and 81\% to 87\% (unseen) relative to SFT, while listen/ignore confusion also tightens. SFT with explicit action tokens outperforms SFT without them (0.903 vs.\ 0.901 seen, 0.907 vs.\ 0.903 unseen), suggesting that the dedicated token vocabulary provides a marginal structural benefit. However, the larger gain comes from GRPO (+2.5 pp over SFT with action tokens), which reinforces correct turn actions directly via action accuracy reward.

\noindent\rule{\linewidth}{0.4pt}

\noindent\textbf{Response Quality (Table~\ref{tab:response_quality}).}
Baselines score low on penalized metrics (e.g., Kimi-Audio's METEOR\,$=$\,0.017 on seen) but recover once conditioned on turns where they do respond, revealing that their main bottleneck is missed responses rather than content quality.
Cocktail-Talker (SFT) already surpasses all baselines on both variants.
GRPO further improves the penalized scores (METEOR: $+$0.011, ROUGE-L: $+$0.015, BERTScore: $+$0.058, SentCos: $+$0.027 on seen), consistent with fewer missed responses inflating the zero-penalty count.
Conditional scores remain roughly stable under GRPO, indicating that GRPO improves turn-taking decisions without degrading response content.
Both variants transfer well to unseen environments with only minor drops.

\noindent\rule{\linewidth}{0.4pt}

\noindent\textbf{Analysis (Fig.~\ref{fig:analysis}).}
We analyze Cocktail-Talker's robustness along three axes: acoustically challenging conditions (a,~b), semantically challenging conditions (c,~d), and long conversational context (e,~f), before examining generalization across conversation environments (g).
On the acoustic side, action accuracy remains strong even at low SNR (a), degrading only slightly from clean audio down to 0--6\,dB mixtures, indicating that the model's turn-taking and response behavior is largely noise-invariant within this range.
As expected, increasing the number of concurrent speakers from three to four (b) consistently lowers both action accuracy and response quality, reflecting the added difficulty of tracking and disambiguating more simultaneous voices in the mixture.
On the semantic side, removing explicit speaker names and roles (c) causes a markedly larger drop than any acoustic factor, since the model must now infer who is being addressed purely from conversational context and voice characteristics rather than from an explicit cue. Casual conversations are similarly harder than assisting conversations (d), because casual dialogue is more open-ended and its turn-taking norms are less determined by an explicit task structure. The model has a harder time judging when a response is appropriate and what it should contain. Examining context length (e,~f), we find a clear non-monotonic pattern: Response quality metrics are strongest with a moderate amount of history, peaking around three to four preceding turns or 30–45 s of audio. Action accuracy remains high for short contexts but declines in several longer-context bins, suggesting that moderate context is generally most useful while longer histories can introduce distracting or redundant information. Finally, across the full set of 28 conversation environments (g), Cocktail-Talker achieves consistently strong performance with only modest variation from environment to environment, and unseen environments, which differ from the training set both acoustically (background sounds) and semantically (conversation topics), show only a slight overall drop relative to seen ones, indicating that the model generalizes well beyond the specific environments it was trained on.

\section{Conclusion and Limitations}

This work explores multi-speaker spoken dialog in noisy social environments, where an assistant selectively participates in a target conversation amid irrelevant speech and noise. We propose \textit{Cocktail-Talker}, which explicitly models and is directly trained to select among three turn actions: respond, listen, and ignore. We further introduce \textit{Cocktail-DialogGen}, an LLM-based pipeline for simulating diverse multi-speaker dialogs and noisy acoustic scenes at scale. \noindent\textbf{Limitations:} Real-world deployment remains challenging. The current system is not streaming and assumes fixed turn boundaries. In addition, non-audio sensors, such as visual and spatial cues, could help the model infer the speaker, addressee, and the appropriate turn action. We leave these directions to future work.

\section*{Acknowledgments}

We used ChatGPT to generate the cartoon illustrations in Fig.~1 and the speaker illustrations in Fig.~2. We also used ChatGPT for text refinement. The original manuscript was written by us, human authors, who reviewed and approved the final text. We used Gemini as a research tool to generate the dialog data, as already described in Section \ref{cocktail-dialoggen}.

We thank the funding from the National Institutes of Health (NIH-NIDCD) and the grant from Marie-Josee and Henry R. Kravis.

\bibliographystyle{IEEEtran}
\bibliography{references}

\end{document}